\documentclass[twocolumn,letterpaper,aps,prd,superscriptaddress,showpacs,nofootinbib,floatfix]{revtex4-1}

\usepackage{graphicx}   
\usepackage{xspace}     

\usepackage{graphicx}
\usepackage{dcolumn}
\usepackage{bm}

\begin{document}


\title{Inclusive cross sections, charge ratio and double-helicity 
asymmetries for $\pi^+$ and $\pi^-$ production in $p$$+$$p$ collisions at 
$\sqrt{s}$=200~GeV}

\newcommand{\abilene}{Abilene Christian University, Abilene, Texas 79699, USA}
\newcommand{\augie}{Department of Physics, Augustana College, Sioux Falls, South Dakota 57197, USA}
\newcommand{\banaras}{Department of Physics, Banaras Hindu University, Varanasi 221005, India}
\newcommand{\barc}{Bhabha Atomic Research Centre, Bombay 400 085, India}
\newcommand{\baruch}{Baruch College, City University of New York, New York, New York, 10010 USA}
\newcommand{\bnlcoll}{Collider-Accelerator Department, Brookhaven National Laboratory, Upton, New York 11973-5000, USA}
\newcommand{\bnlphys}{Physics Department, Brookhaven National Laboratory, Upton, New York 11973-5000, USA}
\newcommand{\caucr}{University of California - Riverside, Riverside, California 92521, USA}
\newcommand{\charlesczech}{Charles University, Ovocn\'{y} trh 5, Praha 1, 116 36, Prague, Czech Republic}
\newcommand{\chonbuk}{Chonbuk National University, Jeonju, 561-756, Korea}
\newcommand{\ciae}{Science and Technology on Nuclear Data Laboratory, China Institute of Atomic Energy, Beijing 102413, People's~Republic~of~China}
\newcommand{\cns}{Center for Nuclear Study, Graduate School of Science, University of Tokyo, 7-3-1 Hongo, Bunkyo, Tokyo 113-0033, Japan}
\newcommand{\colorado}{University of Colorado, Boulder, Colorado 80309, USA}
\newcommand{\columbia}{Columbia University, New York, New York 10027 and Nevis Laboratories, Irvington, New York 10533, USA}
\newcommand{\czechtech}{Czech Technical University, Zikova 4, 166 36 Prague 6, Czech Republic}
\newcommand{\dapnia}{Dapnia, CEA Saclay, F-91191, Gif-sur-Yvette, France}
\newcommand{\elte}{ELTE, E{\"o}tv{\"o}s Lor{\'a}nd University, H-1117 Budapest, P\'azmany P\'eter s\'et\'any 1/A, Hungary}
\newcommand{\ewha}{Ewha Womans University, Seoul 120-750, Korea}
\newcommand{\fsu}{Florida State University, Tallahassee, Florida 32306, USA}
\newcommand{\gsu}{Georgia State University, Atlanta, Georgia 30303, USA}
\newcommand{\hanyang}{Hanyang University, Seoul 133-792, Korea}
\newcommand{\hiroshima}{Hiroshima University, Kagamiyama, Higashi-Hiroshima 739-8526, Japan}
\newcommand{\ihepprot}{IHEP Protvino, State Research Center of Russian Federation, Institute for High Energy Physics, Protvino, 142281, Russia}
\newcommand{\illuiuc}{University of Illinois at Urbana-Champaign, Urbana, Illinois 61801, USA}
\newcommand{\inrras}{Institute for Nuclear Research of the Russian Academy of Sciences, prospekt 60-letiya Oktyabrya 7a, Moscow 117312, Russia}
\newcommand{\instpasczech}{Institute of Physics, Academy of Sciences of the Czech Republic, Na Slovance 2, 182 21 Prague 8, Czech Republic}
\newcommand{\isu}{Iowa State University, Ames, Iowa 50011, USA}
\newcommand{\jaea}{Advanced Science Research Center, Japan Atomic Energy Agency, 2-4 Shirakata Shirane, Tokai-mura, Naka-gun, Ibaraki-ken 319-1195, Japan}
\newcommand{\jyvaskyla}{Helsinki Institute of Physics and University of Jyv{\"a}skyl{\"a}, P.O.Box 35, FI-40014 Jyv{\"a}skyl{\"a}, Finland}
\newcommand{\kek}{KEK, High Energy Accelerator Research Organization, Tsukuba, Ibaraki 305-0801, Japan}
\newcommand{\korea}{Korea University, Seoul, 136-701, Korea}
\newcommand{\kurchatov}{Russian Research Center ``Kurchatov Institute," Moscow, 123098 Russia}
\newcommand{\kyoto}{Kyoto University, Kyoto 606-8502, Japan}
\newcommand{\labllr}{Laboratoire Leprince-Ringuet, Ecole Polytechnique, CNRS-IN2P3, Route de Saclay, F-91128, Palaiseau, France}
\newcommand{\lahorelums}{Physics Department, Lahore University of Management Sciences, Lahore 54792, Pakistan}
\newcommand{\lawllnl}{Lawrence Livermore National Laboratory, Livermore, California 94550, USA}
\newcommand{\losalamos}{Los Alamos National Laboratory, Los Alamos, New Mexico 87545, USA}
\newcommand{\lpc}{LPC, Universit{\'e} Blaise Pascal, CNRS-IN2P3, Clermont-Fd, 63177 Aubiere Cedex, France}
\newcommand{\lund}{Department of Physics, Lund University, Box 118, SE-221 00 Lund, Sweden}
\newcommand{\maryland}{University of Maryland, College Park, Maryland 20742, USA}
\newcommand{\mass}{Department of Physics, University of Massachusetts, Amherst, Massachusetts 01003-9337, USA }
\newcommand{\michigan}{Department of Physics, University of Michigan, Ann Arbor, Michigan 48109-1040, USA}
\newcommand{\muenster}{Institut fur Kernphysik, University of Muenster, D-48149 Muenster, Germany}
\newcommand{\muhlenberg}{Muhlenberg College, Allentown, Pennsylvania 18104-5586, USA}
\newcommand{\myongji}{Myongji University, Yongin, Kyonggido 449-728, Korea}
\newcommand{\nagasaki}{Nagasaki Institute of Applied Science, Nagasaki-shi, Nagasaki 851-0193, Japan}
\newcommand{\newmex}{University of New Mexico, Albuquerque, New Mexico 87131, USA }
\newcommand{\nmsu}{New Mexico State University, Las Cruces, New Mexico 88003, USA}
\newcommand{\ohio}{Department of Physics and Astronomy, Ohio University, Athens, Ohio 45701, USA}
\newcommand{\ornl}{Oak Ridge National Laboratory, Oak Ridge, Tennessee 37831, USA}
\newcommand{\orsay}{IPN-Orsay, Universite Paris Sud, CNRS-IN2P3, BP1, F-91406, Orsay, France}
\newcommand{\peking}{Peking University, Beijing 100871, People's~Republic~of~China}
\newcommand{\pnpi}{PNPI, Petersburg Nuclear Physics Institute, Gatchina, Leningrad Region, 188300, Russia}
\newcommand{\riken}{RIKEN Nishina Center for Accelerator-Based Science, Wako, Saitama 351-0198, Japan}
\newcommand{\rikjrbrc}{RIKEN BNL Research Center, Brookhaven National Laboratory, Upton, New York 11973-5000, USA}
\newcommand{\rikkyo}{Physics Department, Rikkyo University, 3-34-1 Nishi-Ikebukuro, Toshima, Tokyo 171-8501, Japan}
\newcommand{\saopaulo}{Universidade de S{\~a}o Paulo, Instituto de F\'{\i}sica, Caixa Postal 66318, S{\~a}o Paulo CEP05315-970, Brazil}
\newcommand{\stonybrkc}{Chemistry Department, Stony Brook University, SUNY, Stony Brook, New York 11794-3400, USA}
\newcommand{\stonycrkp}{Department of Physics and Astronomy, Stony Brook University, SUNY, Stony Brook, New York 11794-3800, USA}
\newcommand{\tenn}{University of Tennessee, Knoxville, Tennessee 37996, USA}
\newcommand{\titech}{Department of Physics, Tokyo Institute of Technology, Oh-okayama, Meguro, Tokyo 152-8551, Japan}
\newcommand{\tsukuba}{Institute of Physics, University of Tsukuba, Tsukuba, Ibaraki 305, Japan}
\newcommand{\vandy}{Vanderbilt University, Nashville, Tennessee 37235, USA}
\newcommand{\weizmann}{Weizmann Institute, Rehovot 76100, Israel}
\newcommand{\wigner}{Institute for Particle and Nuclear Physics, Wigner Research Centre for Physics, Hungarian Academy of Sciences (Wigner RCP, RMKI) H-1525 Budapest 114, POBox 49, Budapest, Hungary}
\newcommand{\yonsei}{Yonsei University, IPAP, Seoul 120-749, Korea}
\affiliation{\abilene}
\affiliation{\augie}
\affiliation{\banaras}
\affiliation{\barc}
\affiliation{\baruch}
\affiliation{\bnlcoll}
\affiliation{\bnlphys}
\affiliation{\caucr}
\affiliation{\charlesczech}
\affiliation{\chonbuk}
\affiliation{\ciae}
\affiliation{\cns}
\affiliation{\colorado}
\affiliation{\columbia}
\affiliation{\czechtech}
\affiliation{\dapnia}
\affiliation{\elte}
\affiliation{\ewha}
\affiliation{\fsu}
\affiliation{\gsu}
\affiliation{\hanyang}
\affiliation{\hiroshima}
\affiliation{\ihepprot}
\affiliation{\illuiuc}
\affiliation{\inrras}
\affiliation{\instpasczech}
\affiliation{\isu}
\affiliation{\jaea}
\affiliation{\jyvaskyla}
\affiliation{\kek}
\affiliation{\korea}
\affiliation{\kurchatov}
\affiliation{\kyoto}
\affiliation{\labllr}
\affiliation{\lahorelums}
\affiliation{\lawllnl}
\affiliation{\losalamos}
\affiliation{\lpc}
\affiliation{\lund}
\affiliation{\maryland}
\affiliation{\mass}
\affiliation{\michigan}
\affiliation{\muenster}
\affiliation{\muhlenberg}
\affiliation{\myongji}
\affiliation{\nagasaki}
\affiliation{\newmex}
\affiliation{\nmsu}
\affiliation{\ohio}
\affiliation{\ornl}
\affiliation{\orsay}
\affiliation{\peking}
\affiliation{\pnpi}
\affiliation{\riken}
\affiliation{\rikjrbrc}
\affiliation{\rikkyo}
\affiliation{\saopaulo}
\affiliation{\stonybrkc}
\affiliation{\stonycrkp}
\affiliation{\tenn}
\affiliation{\titech}
\affiliation{\tsukuba}
\affiliation{\vandy}
\affiliation{\weizmann}
\affiliation{\wigner}
\affiliation{\yonsei}
\author{A.~Adare} \affiliation{\colorado}
\author{C.~Aidala} \affiliation{\losalamos} \affiliation{\michigan}
\author{N.N.~Ajitanand} \affiliation{\stonybrkc}
\author{Y.~Akiba} \affiliation{\riken} \affiliation{\rikjrbrc}
\author{R.~Akimoto} \affiliation{\cns}
\author{H.~Al-Ta'ani} \affiliation{\nmsu}
\author{J.~Alexander} \affiliation{\stonybrkc}
\author{K.R.~Andrews} \affiliation{\abilene}
\author{A.~Angerami} \affiliation{\columbia}
\author{K.~Aoki} \affiliation{\riken}
% ...
\collaboration{PHENIX Collaboration} \noaffiliation

\date{\today}

\begin{abstract}


We present the midrapidity charged pion invariant cross sections and the 
ratio of $\pi^-$-to-$\pi^+$ production ($5<p_T<13$~GeV/$c$), together with 
the double-helicity asymmetries ($5<p_T<12$~GeV/$c$) in polarized $p$$+$$p$ 
collisions at $\sqrt{s} = 200$~GeV.  The cross section measurements are 
consistent with perturbative calculations in quantum chromodynamics within 
large uncertainties in the calculation due to the choice of factorization, 
renormalization, and fragmentation scales.  However, the theoretical 
calculation of the ratio of $\pi^-$-to-$\pi^+$ production when considering 
these scale uncertainties overestimates the measured value, suggesting 
further investigation of the uncertainties on the charge-separated pion 
fragmentation functions is needed.  Due to cancellations of uncertainties 
in the charge ratio, direct inclusion of these ratio data in future 
parameterizations should improve constraints on the flavor dependence of 
quark fragmentation functions to pions.  By measuring charge-separated 
pion asymmetries, one can gain sensitivity to the sign of $\Delta G$ 
through the opposite sign of the up and down quark helicity distributions 
in conjunction with preferential fragmentation of positive pions from up 
quarks and negative pions from down quarks.  The double-helicity 
asymmetries presented are sensitive to the gluon helicity distribution 
over an $x$ range of $\sim$0.03--0.16.

\end{abstract}

\pacs{Valid PACS appear here}
\maketitle



\section{Introduction}

Quantum chromodynamics (QCD) is well established as the theory of the 
strong interaction, yet it is expressed in terms of quarks and gluons, 
which are confined in color-neutral hadrons and are not observed in 
isolation.  While high-energy scattering of quarks and gluons is 
calculable perturbatively from theory alone, perturbative calculations 
cannot be used without experimental input to determine the quark and gluon 
structure of QCD bound states or the process of hadronization of a 
scattered quark or gluon.  Performing high-energy experimental 
measurements involving hadrons permits the study of nonperturbative 
aspects of QCD supported by the robust and directly calculable framework 
of perturbative QCD (pQCD).  In particular, hadron structure can be 
described in terms of parton distribution functions (PDFs), and parton 
hadronization in terms of fragmentation functions (FFs).  These 
nonperturbative, or long-distance, descriptions factorize from the 
perturbative, or short-distance, partonic hard scattering process, and 
they are universal across different observable reactions.

A great deal about nucleon structure has been learned from experimental 
measurements of inclusive deep-inelastic lepton-nucleon scattering (DIS), 
with complementary knowledge gained from other scattering processes.  
Proton-proton scattering to produce jets, hadrons, or direct photons 
provides access to gluons at leading order in pQCD;  the Drell-Yan process 
of quark-antiquark annihilation to leptons and Drell-Yan-like processes 
such as $W$ and $Z$ boson production enable explicit tagging of antiquarks 
in the nucleon sea.  The dependence of inclusive DIS on the hard momentum 
scale of the scattering, $Q^2$, offers an additional handle on gluon 
distributions for observables measured over a wide range of hard scales, 
and semi-inclusive DIS (SIDIS), in which a produced hadron is observed in 
addition to the scattered lepton, offers another possibility to 
disentangle quark flavor contributions.

Different experimental measurements allow access to different aspects of 
proton structure.  Measurements involving unpolarized protons and a single 
observed hard momentum scale, enabling the framework of pQCD to be 
employed, constrain the unpolarized collinear PDFs; measurements involving 
longitudinally polarized protons and a single hard scale constrain 
helicity-dependent collinear PDFs.  Further, measurements sensitive to 
partonic intrinsic transverse momentum can constrain 
transverse-momentum-dependent (TMD) PDFs, while exclusive measurements 
allow access to spatial distributions of partons within the proton.  All 
of these aspects of the partonic structure of the nucleon can in turn be 
studied in nuclei.  For recent reviews on nucleon structure, 
see~\cite{Aaron:2009aa,Holt:2010vj,Burkardt:2008jw,deFlorian:2011ia,Aidala:2012mv}.

The Relativistic Heavy Ion Collider (RHIC) is an extremely versatile 
hadronic collider, having achieved collisions of ion species from protons 
to uranium over an energy range from a few GeV to several hundred GeV.  
With the ability to control the polarization direction of proton beams 
from 25 to 255~GeV, corresponding to center-of-mass energies of 50 to 
510~GeV, RHIC is excellently suited to study numerous aspects of nucleon 
structure.  One of the main goals of the nucleon structure program at RHIC 
has been to better constrain the helicity PDFs of the proton, given that 
experimental measurements indicate that only 25--35\% of the proton's 
longitudinal spin is carried by the quark 
spins~\cite{Ashman:1987hv,Ashman:1989ig,Adeva:1998vv,Adeva:1998vw,Abe:1998wq}, 
significantly less than the prediction based on the simple parton 
model~\cite{Ellis:1973kp}.

According to the Jaffe-Manohar sum rule for the spin of the 
proton~\cite{Jaffe:1989jz}, the proton's longitudinal spin can be 
decomposed as $\frac{1}{2}=\frac{1}{2}\Delta \Sigma + \Delta G + L_{q}+ 
L_{g}$, where $\frac{1}{2}\Delta \Sigma$ is the net quark and antiquark 
spin, $\Delta G$ is the net gluon spin, and $L_{q}+L_{g}$ is the orbital 
angular momentum of partons. Within the Jaffe-Manohar decomposition, it is 
known how to compare $\Delta \Sigma$ and $\Delta G$ to experimental 
measurements as they are gauge invariant. Particularly, $\Delta G$ has a 
physical interpretation as gluon spin in the light cone gauge. For a 
recent review of helicity sum rules and the interpretations of their 
components, see~\cite{Leader:2013jra,Wakamatsu:2014zza}.

RHIC measurements of the double-longitudinal spin asymmetry in the 
production of hadrons~\cite{Adare:2008aa, Adare:2008qb} and 
jets~\cite{Abelev:2007vt}, as well as single-longitudinal spin asymmetry 
measurements in the production of $W$ bosons~\cite{Adare:2010xa, 
Aggarwal:2010vc}, have already been used in helicity PDF 
fits~\cite{deFlorian:2008mr, deFlorian:2009vb, Aschenauer:2013woa, 
Nocera:2014gqa, deFlorian:2014yva}.  Furthermore, large transverse 
single-spin asymmetries in forward particle production, sensitive to 
spin-momentum correlations in QCD, have been measured at RHIC in 
transversely polarized proton 
collisions~\cite{Adams:2003fx,Adamczyk:2012xd,Arsene:2008aa,Lee:2009ck,Adare:2013ekj}.

Using hadrons in the final state to study nucleon structure, i.e.~in the 
processes of $p$$+$$p$ to hadrons or SIDIS, requires knowledge of FFs.  
Fragmentation functions are in a sense complementary to PDFs.  While PDFs 
describe the behavior of colored quarks and gluons bound in a colorless 
hadron, FFs describe a colored quark or gluon transitioning into one or 
more colorless hadrons.  As in the case of PDFs, different experimental 
measurements can provide a variety of information on FFs.  Much has been 
learned from the process of electron-positron annihilation to hadrons, 
while $p$$+$$p$ collisions are especially useful in constraining gluon FFs, 
and SIDIS data can provide information on quark flavor.  Hadron production 
in $p$$+$$p$ collisions offers additional flavor sensitivity, and the $Q^2$ 
evolution of hadrons produced in $e^+e^-$ provides an additional handle on 
gluons.  Also similar to the case of PDFs, different observables allow the 
study of unpolarized collinear FFs, unpolarized TMD FFs, and various 
spin-momentum correlations in hadronization through yet other TMD FF 
measurements.  Recently, measurements in $e^+e^-$ of two hadrons 
fragmenting from the same parton have opened up the exploration of 
dihadron FFs~\cite{Vossen:2011fk}.  Cross section measurements of 
inclusive hadron production at 
RHIC~\cite{Adler:2003pb,Abelev:2006cs,Adams:2006nd,Adams:2006uz,Arsene:2007jd,Adare:2010cy} 
have already been used to help constrain FFs for charged and neutral pions 
and kaons, protons, and 
lambdas~\cite{deFlorian:2007aj,deFlorian:2007hc,Albino:2008fy,Aidala:2010bn,Epele:2012vg}.  
Measurements of back-to-back direct photon and charged hadron production 
to study hadronization in $p$$+$$p$ collisions at RHIC have also been 
performed~\cite{Adare:2010yw}.  For a review of FFs, 
see~\cite{Albino:2008gy}.

This paper presents charged-pion cross section measurements in 
polarization-averaged $p$+$p$ collisions and double-helicity 
asymmetry measurements in charged pion production in longitudinally 
polarized proton collisions at $\sqrt{s}=200$~GeV.  The cross section 
measurements provide information on charge-separated collinear FFs for 
pions; the spin asymmetry measurements are sensitive to the collinear 
helicity distributions in the proton.  Improved knowledge of FFs can lead 
to improved knowledge of PDFs, and vice versa, in a continuous, iterative 
process of increasingly refined constraints on nonperturbative aspects of 
QCD.  In Sec.~\ref{sec:experiment} we describe the PHENIX experiment; 
Sec.~\ref{sec:analysis} discusses the analysis; Sec.~\ref{sec:results} 
presents the results, which are further discussed in 
Sec.~\ref{sec:discussion}; and Sec.~\ref{sec:summary} summarizes our 
findings.

\section{The PHENIX Experiment}
\label{sec:experiment}

A large set of data was taken in 2009 from polarized $p$$+$$p$ collisions at 
$\sqrt{s}$~=~200~GeV, and an integrated luminosity of 11.8~pb$^{-1}$ was 
analyzed for charged pion production at midrapidity. The PHENIX detector 
is configured with two central arm spectrometers each covering 90$^\circ$ 
in azimuth, $|\eta|<0.35$ in pseudorapidity, two forward muon 
spectrometers (not discussed further) and two sets of detectors to 
determine collision parameters.

\begin{figure}[htb]
\includegraphics[width=1.0\linewidth]{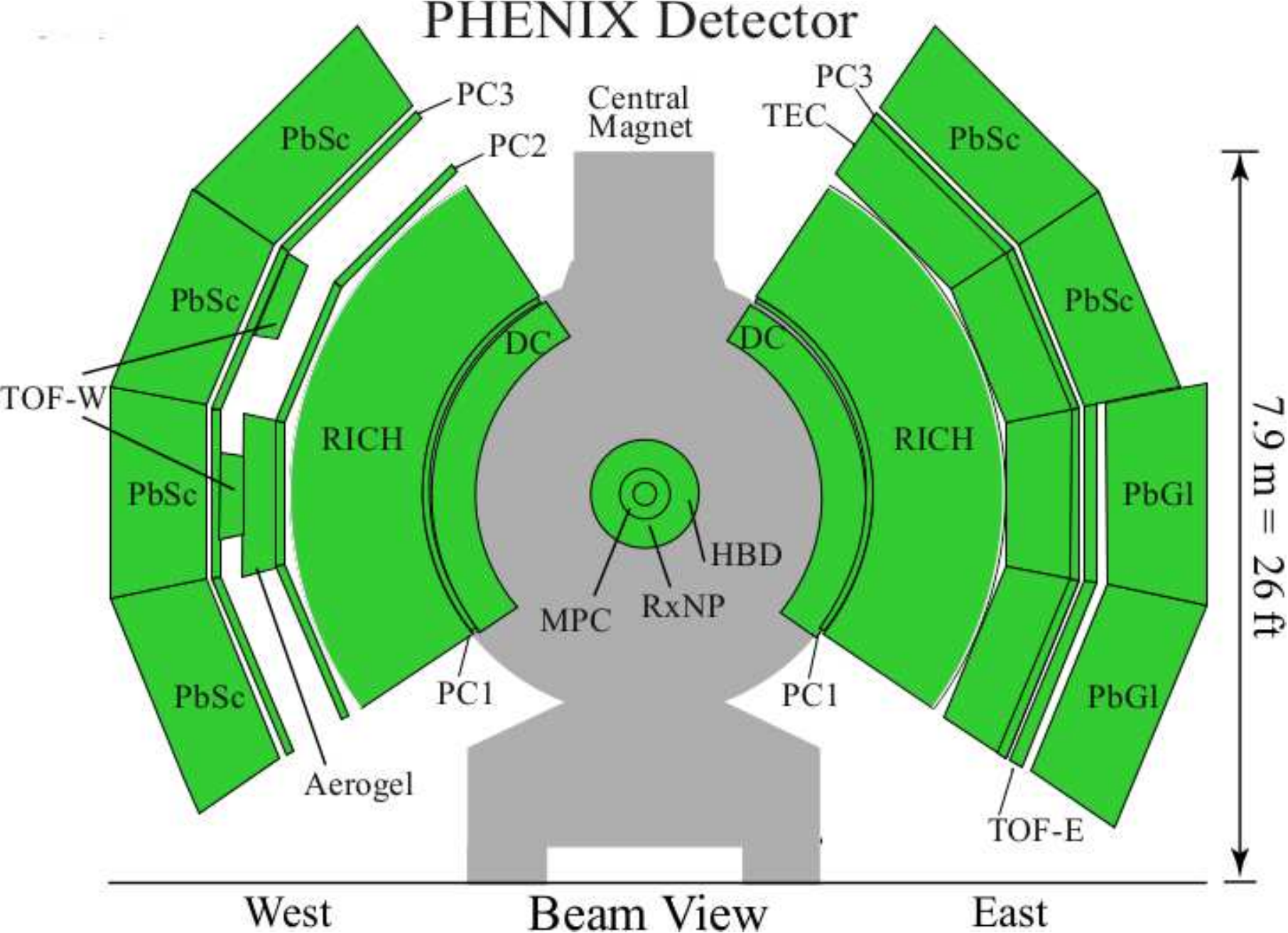}
\caption{\label{phenix} (color online)
Cross section view along the beam line of the PHENIX detector, showing 
the detectors composing the central arms described in the text, as well 
as the HBD added in 2009.}
\end{figure}

The central arms comprise several tracking layers composed of a drift 
chamber (DC) and pad chambers (PC), a ring-imaging \v{C}erenkov (RICH) 
detector for particle identification, and an electromagnetic calorimeter 
(EMCal).  In 2009, an additional \v{C}erenkov detector, the hadron-blind 
detector (HBD), was added to reduce the combinatorial background in 
low-mass dilepton measurements in heavy ion collisions; this detector can 
also be used for charged pion identification.  Results on single electrons 
from heavy flavor decays measured using the HBD have already been 
published~\cite{Adare:2012vv}.  A layout of the central arms is shown in 
Fig.~\ref{phenix}.

\subsection{Collision Detectors}

Two Beam-Beam Counters (BBCs) \cite{Allen:2003zt} placed at 144~cm on 
either side of the interaction point (IP) of PHENIX along the beam axis 
were used for event start timing, triggering, luminosity monitoring and 
determination of the location of the collision vertex along the beam axis. 
The BBCs are built with 64 photomultiplier tubes, each coupled to a quartz 
\v{C}erenkov radiator. An offline cut of 30~cm in the distribution of the 
reconstructed collision point was used for event selection.

Additionally, two Zero-Degree Calorimeters (ZDCs) are located 18~m from 
the IP along the beam axis, after the RHIC beam bending magnets.  Each ZDC 
comprises three sections of hadron calorimeter composed of optical 
fibers for \v{C}erenkov sampling between layers of tungsten absorber.  
Between the first and second sections is placed a Shower Maximum Detector 
(SMD), which comprises vertical and horizontal scintillator strips, and 
is designed to determine the shower maximum of hadronic (primarily 
neutron) showers.  The ZDCs are used as a comparison luminosity monitor 
and, with the SMD, to ensure transverse components of the beam 
polarization are small.  Any remaining transverse components of the beam 
polarization can be determined by the transverse single-spin asymmetry 
observed for forward neutron production at RHIC~\cite{Adare:2012vw}.

\subsection{Magnetic Field and Tracking Detectors}

The central magnet~\cite{Aronson:2003zn} in PHENIX supplies an axial field 
for momentum measurements of charged particles.  It comprises two coils, 
the inner and outer, which can be run independently.  To have a 
field free region in the HBD volume in 2009, the coils were run in an 
opposing `$+-$' configuration.  In this configuration, the field strength 
is near zero for $0<R<50$~cm, corresponding to the HBD region, and then 
increases to a peak of 0.35~T around 1~m, with a total magnetic field 
integral of 0.43 Tm.  It is designed such that the field strength is near 
zero at 2~m, where the first tracking layers are located, and is required 
to be less than 100~Gauss in the region of the RICH photomultiplier tubes.

The primary tracking device to determine momentum is the 
DC~\cite{Adcox:2003zp}, located radially between 2 and 2.5~m from the 
$z$-axis in a region of low magnetic field strength. The transverse 
component of the track momentum with respect to the beam particle 
direction is determined based on the angle between the reconstructed 
track, after bending in the magnetic field, and the straight line from the 
$z$-axis to the track midpoint in the DC.  The first layer of the PC 
detector~\cite{Adcox:2003zp} (PC1) sits behind the DC radially, and is 
used along with the reconstructed collision vertex to determine the 
component of the momentum parallel to the $z$-axis.  A third layer of the 
PC (PC3) is located radially directly in front of the EMCal, and is used 
to ensure track quality.

\subsection{Electromagnetic Calorimeter}

The EMCal \cite{Aphecetche:2003zr} was used for energy measurement and 
triggering. Six of the eight EMCal sectors are constructed from 
lead-scintillator (PbSc) towers in a sampling configuration while the 
remaining two sectors are made of lead-glass (PbGl) towers.  The PbSc 
(PbGl) EMCal corresponds to 0.85 (1.05) nuclear interaction lengths.

A high energy trigger, primarily designed for photons, is implemented in 
the EMCal, and was used to select high momentum charged pion candidates.  
Events were triggered by requiring an EMCal cluster with an energy 
threshold of 1.4~GeV in addition to a BBC coincidence trigger.

\subsection{Charged Pion Identification}

In 2009, PHENIX had two \v{C}erenkov based charged particle identification 
detectors:  the RICH and the HBD.  While both were primarily designed to 
identify electrons, they can also be used to identify pions above 
$p_T\sim$~4.7~GeV/$c$.

The RICH \cite{Aizawa:2003zq}, which uses CO$_{2}$ as a radiator, allows 
for the identification of charged pions above $p_T\sim$~4.7~GeV/$c$. The 
\v{C}erenkov light is collected by a photomultiplier array on a plane 
outside of the tracking acceptance after reflection by a pair of focusing 
spherical mirrors.

The HBD \cite{Anderson:2011jw} is a gas electron multiplier (GEM) based 
\v{C}erenkov detector. Located at $\sim$5~cm from the beam pipe with the 
windowless CF$_4$ radiator extending to $\sim$60 cm~in the radial 
direction, it covered a pseudorapidity range of $|\eta|$$<$0.45.  The 
$p_{T}$ thresholds of \v{C}erenkov radiation for electrons, pions, and 
kaons in CF$_4$ are $\sim 1, 4$, and 14~GeV/$c$, respectively.  The 
\v{C}erenkov photons generate photoelectrons on a CsI photocathode layer 
on the first GEM foil which are subsequently amplified as they traverse 
the GEM holes and collected in readout pads.  Because electrons produced in 
the avalanche can be distributed over more than one readout pad, adjacent 
pads with charge above the pedestal are grouped together to form a 
cluster. The total cluster charge is used as a variable for particle 
identification. In addition to \v{C}erenkov photons, scintillation photons 
can also be generated by charged particles moving inside the radiator. The 
mean number of photons created by scintillation per charged particle 
($\sim$1) is much smaller than that created through \v{C}erenkov 
radiation. At $p_T>5$~GeV/$c$, electrons produce a mean of 
20~photoelectrons while the yield for pions has a $p_T$ dependence due to 
the high threshold momentum.

\section{Data Analysis}
\label{sec:analysis}

Charged pion candidates are selected based on track quality and particle 
identification cuts in the \v{C}erenkov detectors.  Each candidate is 
required to have a high quality, unique track defined in the DC and an 
associated PC1 hit.  Further, the track is required to match with hits in 
the PC3 and EMCal.  Candidate tracks are required to be associated with an 
EMCal trigger.  For the RICH, we define the variable, $n_1$, as the total 
number of photomultipliers that fired within a radius of 11~cm around the 
projected track position. For charged pion candidates, we require $n_1>0$.

\begin{figure}[htb]
\includegraphics[width=1.0\linewidth]{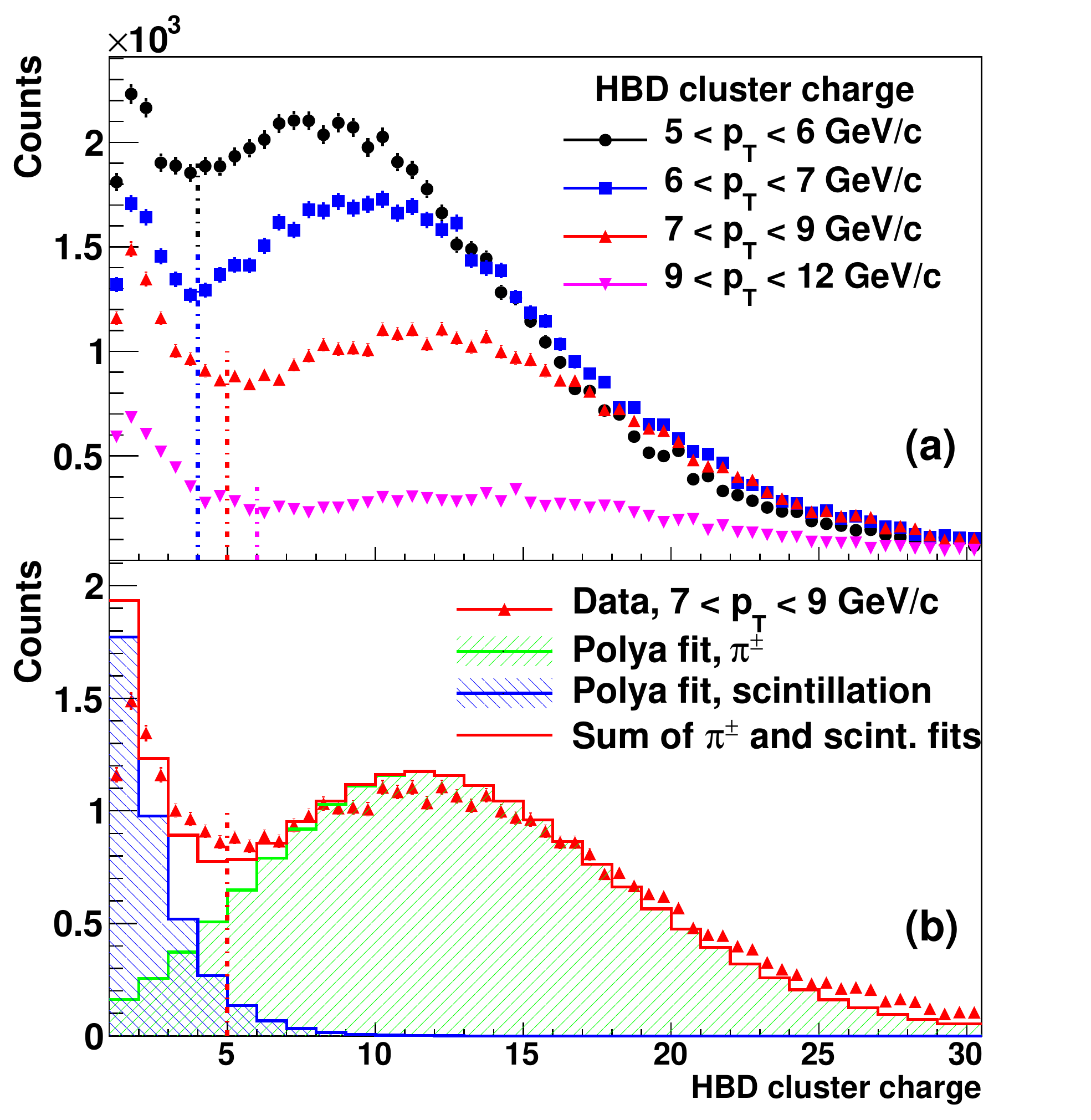}
\caption{\label{hbdq} (color online)
The HBD cluster charge distribution, in units of photoelectrons.  Cuts 
applied on the cluster charge for the four $p_T$ bins are shown as dashed 
lines. Note that the cluster charge cut for tracks reconstructed with 5--6 
and 6--7~GeV/$c$ was the same.}
\end{figure}

A further cut is applied based on the HBD cluster charge. 
Figure~\ref{hbdq} shows the HBD cluster charge distribution of $\pi^{\pm}$ 
candidates after applying all cuts other than the HBD cluster charge cut.  
The cluster charge distributions for four $p_T$ bins are shown in 
Fig.~\ref{hbdq}(a). The peak on the right for each $p_T$ bin is from 
$\pi^{\pm}$. The mean number of photoelectrons generated before avalanche 
is extracted for each peak by fitting the charge distribution with a 
statistical model distribution known as the folded-Polya probability 
distribution. The analytic form of the folded Polya distribution is 
derived from the Polya distribution by convolution with an avalanche 
process model. Fitting results consistently describe the rising mean 
number of \v{C}erenkov photons with $p_T$. The secondary peak found on the 
left comes from scintillations in the HBD. A sum of two folded-Polya 
functions with independent weights was used as a fit function for the 
whole charge distribution. An example of fitted cluster charge 
distribution is shown in Fig.~\ref{hbdq}(b).

Tracks accidentally associated with scintillation charge on the HBD mainly 
comprise electrons from photon conversions and from decays of long lived 
hadrons outside the HBD.  As the first tracking detectors sit outside of 
the magnetic field region, the momentum of such electrons created far from 
the collision point will be incorrectly reconstructed, and then may be 
incorrectly associated with a track-matching cluster charge in the HBD. 
Applying $p_T$-dependent cuts on the cluster charge (vertical lines in 
Fig.~\ref{hbdq}) resulted in effective removal of these incorrectly 
reconstructed electrons. Residual background after applying these cuts is 
estimated by taking the count ratio of the scintillation to the pion 
contribution above the charge cut. Compared with the analysis without the 
HBD, inclusion of the HBD in particle identification resulted in 
improvement in this background source from $\sim 30$\% to $<$2\% averaged 
over the $p_T$ range.

The only remaining background that could pass the requirement of a 
\v{C}erenkov cluster in the HBD is electrons created prior to the HBD 
back plane, such as from decays of neutral pions (dominant), $K_{e3}$ 
electrons ($K^{\pm}/K_{L}\rightarrow\pi^0e^{\pm}\nu$), heavy flavor and 
vector mesons. As such electrons are generated prior to traversing the 
magnetic field region, their reconstructed momentum is generally correct.  
However, these contributions are suppressed due to small cross sections 
and small branching ratios in the case of heavy flavor and vector mesons, 
and small conversion or decay probabilities inside the HBD for electrons 
from $\pi^0$ and $K_{e3}$.  In the case of decay electrons, decay 
kinematics further reduces their rate because the $p_T$ distribution of 
decay electrons is softer than that of the parent particles. The rate of 
electrons from the largest contributors (conversion of $\pi^0$-decay 
photons and $\pi^0$ Dalitz decay) evaluated by these considerations of 
cross sections and branching ratios amount to $\lesssim 2.7$\% of the rate 
of signal events, while other sources combined are less than 1\%.

\begin{figure}[t]
\includegraphics[width=1.0\linewidth]{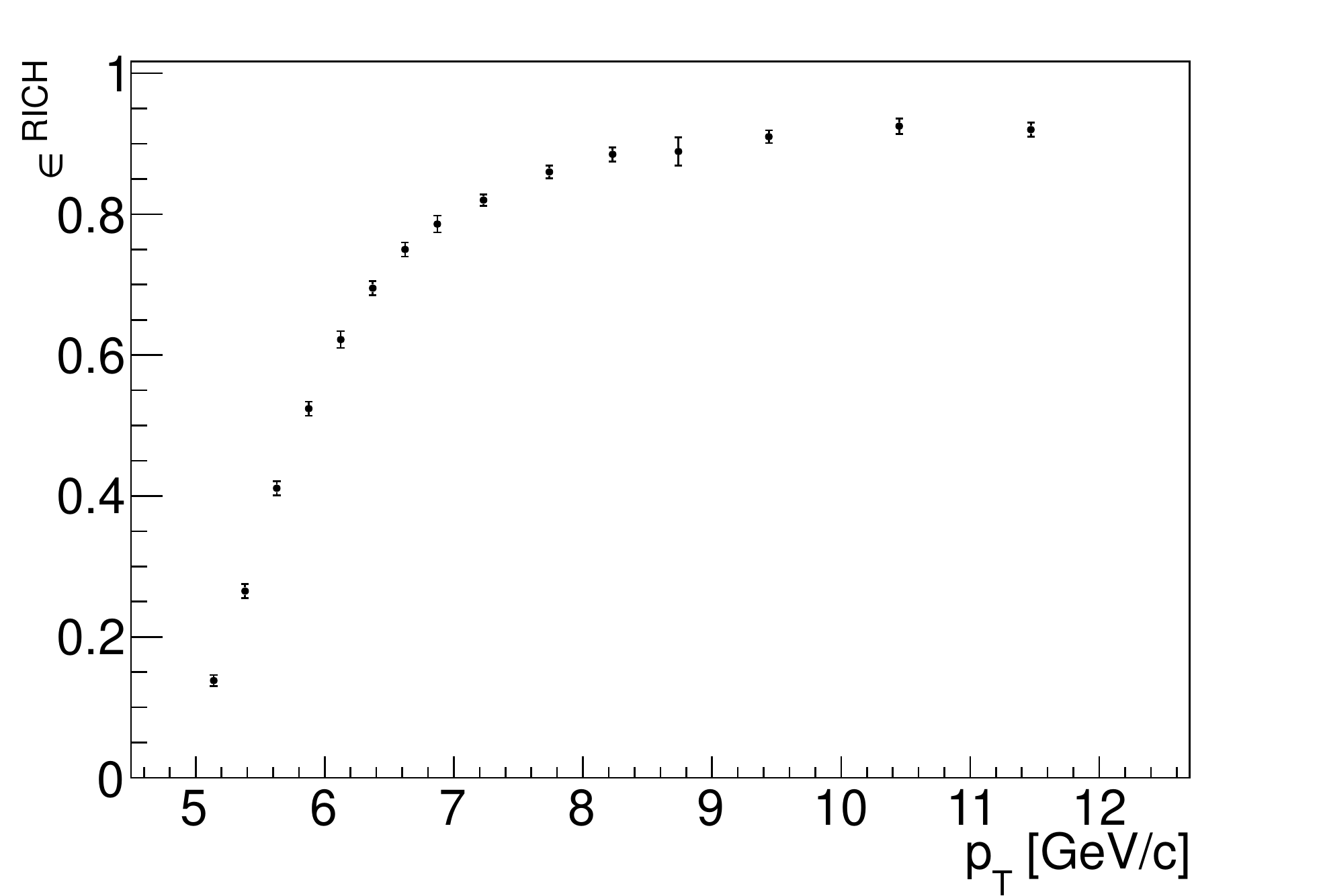}
\caption{\label{richeff} 
The RICH efficiency as a function of pion $p_T$ 
above the pion \v{C}erenkov threshold in the RICH.}
\end{figure}

For a cross section measurement, the track reconstruction efficiencies 
must be determined.  For this measurement, we have separated the geometric 
acceptance, which is determined from MC simulations, and detector 
efficiencies, which are determined from data.  The systematic uncertainty 
of the geometric acceptance was determined to be 2\% by varying the 
boundary of the active detection area in the simulation.

For determination of detector efficiencies, we calculate the survival 
probability, when requiring a hit in the active area of a detector, of a 
sample of clean $\pi^{\pm}$ tracks that leave a signal in all other 
detectors.  This sample was acquired by using tighter cuts to enhance the 
signal fraction. First, dead areas were identified by comparing hit and 
projected track distributions and masked. This was done to obtain 
pure detector efficiencies, not convoluted with dead area effect. Data 
were then divided into groups to account for the variation of efficiency 
with time.  Within each group, the weighted average of the fill-by-fill 
efficiency was used as an effective efficiency, with the RMS assigned as 
the systematic uncertainty (2.8\% for the DC, 4.5\% for the PC and 1.6\% 
for the HBD).

For the RICH, using the method described in the previous paragraph is 
particularly advantageous as it allows measuring the sharply rising 
efficiency near the \v{C}erenkov threshold in a model-independent way. 
Model based MC simulations do not well describe this efficiency 
turn-on. Figure~\ref{richeff} shows the measured RICH efficiency as a 
function of $p_T$ showing a clear rise above the RICH \v{C}erenkov 
threshold for pions. The systematic uncertainties on the RICH efficiency 
were determined by comparing with a Fermi function used to describe the 
turn on, and varied from 12\% at 5~GeV$/c$ to $\sim$1\% at high $p_T$.


The momentum smearing effect caused by finite resolution of measured $p_T$ 
was corrected by unfolding the measured cross section using MC 
simulation. The unfolded $p_T$ distribution corresponds to the true $p_T$ 
distribution given that the $p_T$ was affected by the resolution function 
of PHENIX tracking system, found to be $\sqrt{(1.74)^2 + (1.48 \times p_T 
[\textrm{GeV}/c])^2}$\%, during measurements. The hardening factors of the 
$p_T$ spectra, determined $p_T$ bin by $p_T$ bin by taking the ratio 
between the fitted curves of the unfolded $p_T$ distribution and the 
measured cross section, were negligible below 8~GeV$/c$ but become 
significant ($\sim 16$\%) at the highest $p_T$ bin due to the sufficiently 
large $p_T$ resolution. The systematic uncertainty from momentum smearing 
was estimated to be ~1\% by comparing the fitted curves of measured cross 
section and the re-smeared spectra of unfolded results. The corrections on 
the cross section measurements attributed to the momentum smearing effect 
are large due to the rapidly falling shape of the cross section. However, 
the impact of momentum smearing on the $A_{LL}$ asymmetry measurements can 
be ignored as the asymmetries vary much more slowly as a function of 
$p_T$.

\begin{figure}[htb]
\includegraphics[width=1.0\linewidth]{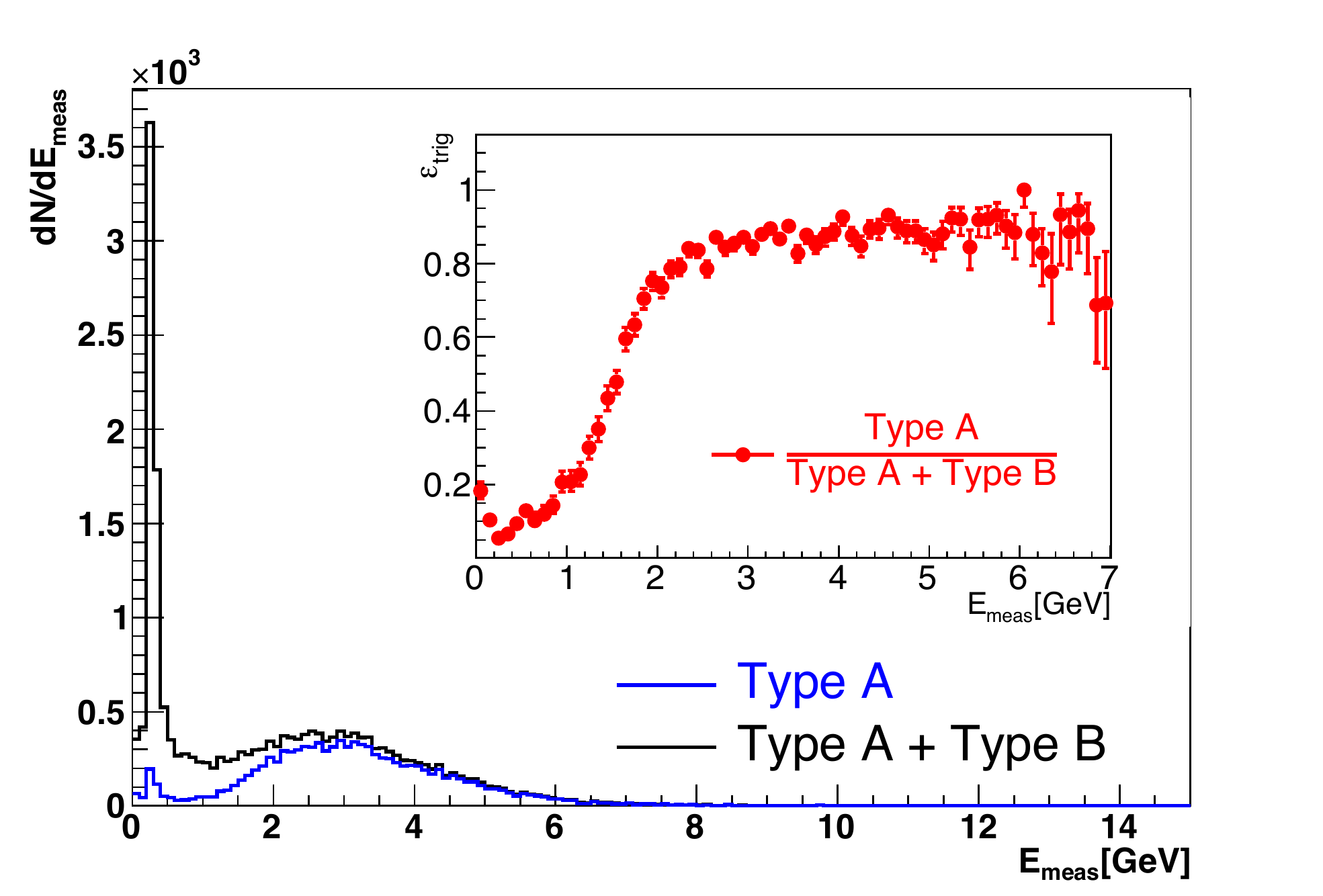}
\caption{\label{trigeff}  (color online)
The energy spectra of (black) all charged pions (type A+type B) and (blue) 
those that fire the EMCal trigger (type A).  The inset shows the resulting 
EMCal trigger efficiency for charged pions as a function of deposited 
energy.}
\end{figure}

The trigger efficiency is an important absolute normalization factor in 
the measurement of invariant differential cross sections. As PHENIX does 
not have a dedicated charged hadron trigger, a new method was developed to 
determine the EMCal trigger efficiency for events containing charged 
pions. This method exploits the event structure to select a high 
statistics subset of the data from which the EMCal trigger bias is 
removed. This subset is composed of events containing a high $p_T$ charged 
pion in addition to one or more spatially separated particles which fired 
the EMCal trigger.  To avoid possible bias from a trigger associated with 
a nearby particle, a minimum energy cut of 0.2~GeV was applied.  These 
events are divided into two types:  type A where the pion does fire the 
trigger and type B otherwise.  Figure~\ref{trigeff} shows the spectra for 
all pions in black and type A--those that fire the trigger--in blue. We 
define the trigger efficiency as the ratio of type A event counts to the 
sum of type A and type B event counts (i.e., all events) above the minimum 
energy cutoff. The inset in Fig.~\ref{trigeff} shows this ratio as a 
function of the deposited energy in the EMCal for charged pions of 
5$<p_T<$13~GeV/$c$. The average trigger efficiency for all energies above 
the cutoff was also calculated as a function of $p_T$, and shows no $p_T$ 
dependence. A constant fit over the whole $p_T$ range yields value of 
0.497$\pm$0.7\% for the trigger efficiency with a $\chi^2$/d.o.f.=0.88.

The full list of systematic uncertainties associated with the correction 
factors discussed so far is summarized in Table~\ref{systematics}. The 
systematic uncertainties reported for the ratio of the cross sections of 
$\pi^+$ to $\pi^-$ are calculated as the quadratic sum of those 
uncertainties that do not cancel out between the two measurements. A 
systematic uncertainty is considered to cancel between the two 
measurements if a change in the underlying cause modifies the cross 
sections of $\pi^-$ and $\pi^+$ by the same multiplicative factor. This is 
the case for instance with all the uncertainties related to the efficiency 
of individual detectors. A misdetermination of these efficiencies would 
affect the $\pi^+$ and $\pi^-$ cross sections identically, hence its 
effect would not be visible in the ratio. However, the geometric 
acceptance uncertainties do not cancel for the two charges. This is 
because the way dead areas are seen by tracks of opposite signs are 
different, and hence, a change in detector dead area configuration will 
not affect the geometrical acceptances in the same way for $\pi^+$ and 
$\pi^-$. The same is true for the $p_T$ smearing correction uncertainty. 
These are the only two uncertainties considered in the ratio measurement.

\begin{table}[tbh]
\caption{Summary of systematic uncertainties for each $p_T$ bin (in \%) 
The $\epsilon^{\textrm{reco}}_{\textrm{DET}}$ stands for the 
reconstruction efficiency of each detector used for this analysis, while 
the remaining three uncertainties are related to geometrical acceptance 
correction $\epsilon^{\textrm{geo}}_{\textrm{acc}}$, trigger efficiency 
correction $\epsilon^{\textrm{trig.}}_{\textrm{eff.}}$, and $p_T$ smearing 
correction $\epsilon^{p_T}_{\textrm{smear}}$ }
\begin{ruledtabular} \begin{tabular}{c c c c c c c c}
&&&&&&&\\
 $p_T$ bin &$\epsilon^{\textrm{reco}}_{\textrm{\rm DC}}$& $\epsilon^{\textrm{reco}}_{\textrm{HBD}}$  &	$\epsilon^{\textrm{reco}}_{\textrm{PC3}}$& $\epsilon^{\textrm{reco}}_{\textrm{\rm RICH}}$ & $\epsilon^{\textrm{geo.}}_{{\rm acc.}}$& $\epsilon^{\textrm{trig.}}_{\textrm{ eff.}}$ & $\epsilon^{p_T}_{\textrm{smear}}$\\
(GeV/$c$)&&&&&&&\\
\hline
5--6 &2.8&1.6&4.5& 12.1&2.0& 1.4 &1.0\\
6--7 &2.8&1.6&4.5&2.6&2.0&1.2&1.0\\
7--8 &2.8&1.6&4.5&0.9&2.0&1.6&1.0\\
8--9 &2.8&1.6&4.5&0.6&2.0&2.2&1.0\\
9--11 &2.8&1.6&4.5&0.5&2.0&2.3&1.0\\
11--13&2.8&1.6&4.5&0.5&2.0&5.1&1.0\\
\end{tabular} \end{ruledtabular} 
\label{systematics}
\end{table}

\begin{figure*}[htb]
\includegraphics[width=0.998\linewidth]{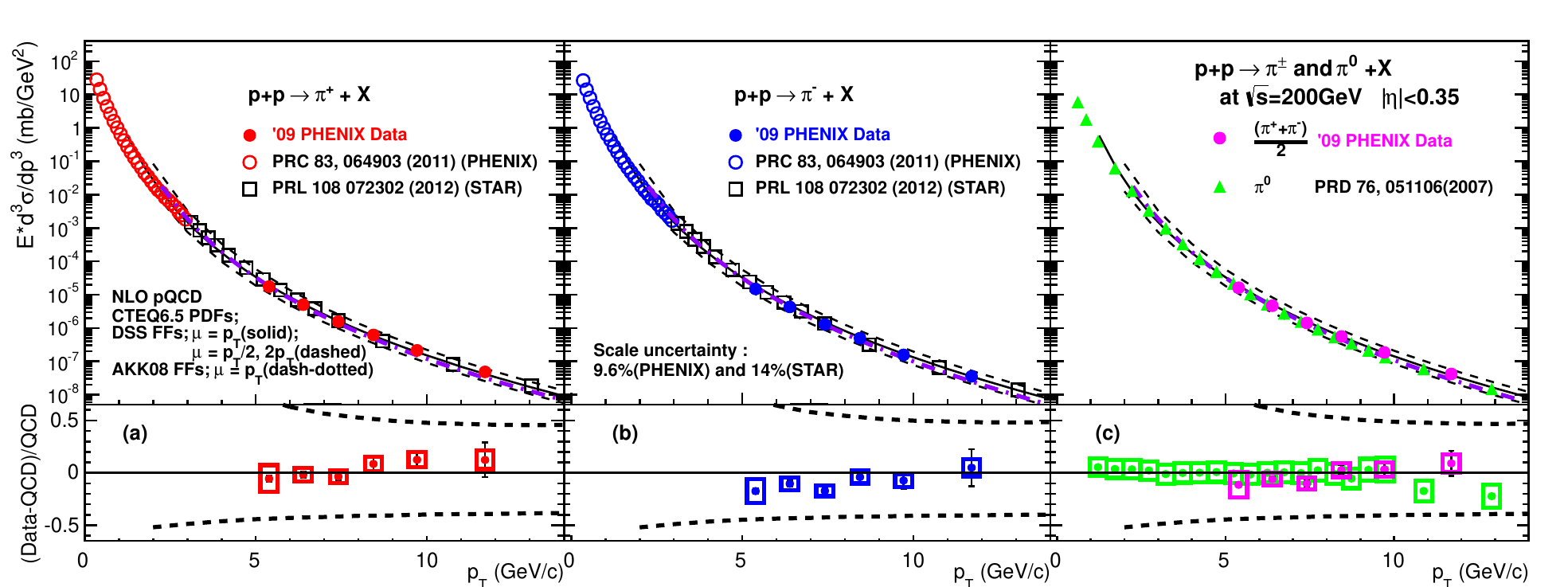}
\caption{\label{xsecpipm}   (color online)
Invariant cross sections for (a) $\pi^+$ and (b) $\pi^-$ with pQCD 
predictions using the DSS~\cite{deFlorian:2007aj} and AKK08~\cite{ 
Albino:2008fy} FFs. Top panel: PHENIX~\cite{Adare:2011vy} and 
STAR~\cite{Agakishiev:2011dc} results are also compared. Bottom: 
systematic (boxes) and statistical (bars) uncertainties are shown with 
relative difference between data and prediction. (c) Comparison of 
averaged charged pion cross section and $\pi^0$ cross section by 
PHENIX~\cite{Adare:2007dg}. Bottom panel: data-theory comparisons.}
\end{figure*}

\section{Results}
\label{sec:results}
\subsection{Cross Sections}
\label{sec:xsecresults}

The cross section measurements for $\pi^+$ and $\pi^-$ hadrons are shown 
in Fig.~\ref{xsecpipm}(a) and (b), respectively, along with previously 
published PHENIX results~\cite{Adare:2011vy} for low-$p_T$ charged pions 
measured with the PHENIX Time of Flight detector~\cite{Aizawa:2003zq}, 
which can identify pions for $p_T<3$~GeV/$c$.  There is an overall scale 
uncertainty from the determination of the absolute luminosity sampled of 
9.6\%, correlated between all PHENIX results.  This absolute normalization 
uncertainty derives from the uncertainty on the inelastic $p$$+$$p$ cross 
section sampled by the BBC trigger, which is found to be 23.0$\pm$2.2~mb 
based on van der Meer scan results~\cite{Adler:2003pb} corrected for 
year-to-year variations in the BBC performance.  Results from the STAR 
experiment~\cite{Agakishiev:2011dc} are also plotted and are consistent 
with the new PHENIX data.

In Fig.~\ref{xsecpipm}(c), the charge-averaged pion cross section is shown 
along with the previously published PHENIX neutral pion cross 
section~\cite{Adare:2007dg}.  The charged and neutral pion measurement are 
found to be in good agreement with each other, as can be seen in the lower 
panel where both are compared to pQCD calculations~\cite{marco}.  The 
comparison of the measurements to the theoretical calculations is 
discussed further in Sec.~\ref{sec:discussion} below.

\begin{table*}[tbh]
\caption{Invariant cross section for $\pi^+$ and $\pi^-$ hadrons, as well 
as the statistical and systematic uncertainties.  In addition, there is an 
absolute scale uncertainty of 9.6\%.}
\begin{ruledtabular} \begin{tabular}{cccccccccc}
&
&& \multicolumn{3}{c}{$\pi^+$}
&& \multicolumn{3}{c}{$\pi^-$} \\
  $p_T$ bin
 & $\langle p_T\rangle$
&& $E*\frac{d^3\sigma}{dp^3}$
& STAT & SYST
 &&  $E*\frac{d^3\sigma}{dp^3}$
& STAT & SYST  \\
(GeV/$c$) & (GeV/$c$)
&& $(mb/GeV^2)$ & &
&& $(mb/GeV^2)$ & & \\
\hline
5--6      &  5.39     &&  1.75$\times10^{-5}$    &  0.05$\times10^{-5}$       &  0.24$\times10^{-5}$      &&  1.49$\times10^{-5}$    &  0.04$\times10^{-5}$    &  0.20$\times10^{-5}$  \\
6--7      &  6.39     &&  5.01$\times10^{-6}$    &  0.15$\times10^{-6}$       &  0.33$\times10^{-6}$      &&  4.30$\times10^{-6}$    &  0.13$\times10^{-6}$    &  0.29$\times10^{-6}$  \\
7--8      &  7.41     &&  1.56$\times10^{-6}$  &  0.07$\times10^{-6}$    &  0.10$\times10^{-6}$    &&  1.283$\times10^{-6}$  &  0.060$\times10^{-6}$  &  0.080$\times10^{-6}$  \\
8--9      &  8.44     &&  6.19$\times10^{-7}$    &  0.39$\times10^{-7}$       &  0.40$\times10^{-7}$      &&  4.94$\times10^{-7}$    &  0.35$\times10^{-7}$    &  0.32$\times10^{-7}$ \\
9--11    &  9.71     &&  2.14$\times10^{-7}$    &  0.16$\times10^{-7}$      &  0.14$\times10^{-7}$       &&  1.57$\times10^{-7}$    &  0.13$\times10^{-7}$    &  0.10$\times10^{-7}$  \\
11--13  &  11.70  &&  4.83$\times10^{-8}$    &  0.71$\times10^{-8}$       &  0.38$\times10^{-8}$      &&  3.57$\times10^{-8}$    &  0.60$\times10^{-8}$     &  0.28$\times10^{-8}$  \\
\end{tabular} \end{ruledtabular}
\label{t:pi0_final_datatable}
\end{table*}

\subsection{Cross Section Ratio}
\label{sec:xsectratio}

\begin{figure}[htb]
\includegraphics[width=1.0\linewidth]{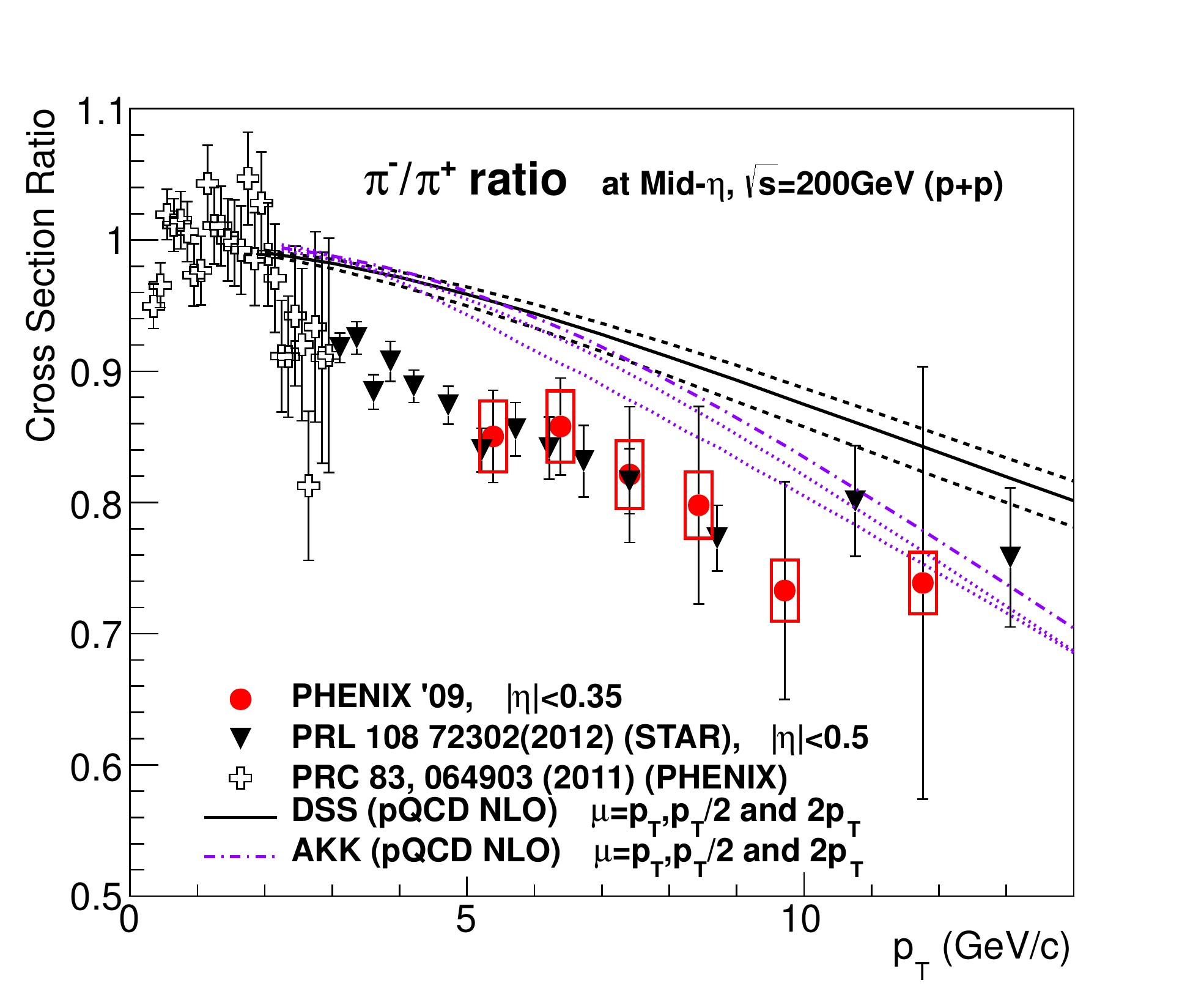}
\caption{\label{ratio} (color online) 
The ratio of the pion production cross sections for the two charges. 
Systematic uncertainties are shown in red boxes. Also shown are the PHENIX 
low $p_T$ results and the STAR high $p_T$ results.}

\end{figure}

The $\pi^-$-to-$\pi^+$ cross section ratio from this analysis is shown in 
Fig.~\ref{ratio} together with the measurements from 
PHENIX~\cite{Adare:2011vy} at lower $p_T$ and from 
STAR~\cite{Agakishiev:2011dc} in a similar $p_T$ range. The results from 
the three measurements are compatible. The experimental data is compared 
to pQCD calculations based on DSS~\cite{deFlorian:2007aj,Epele:2012vg} and 
AKK~\cite{Albino:2008fy} FFs. The calculated ratio values for theory scale 
choices of $\mu = p_T$, $\mu = 0.5p_T$, and $\mu = 2p_T$ are additionally 
shown.

\subsection{Helicity Asymmetries}
\label{sec:allresults}

Hard processes in polarized $p$$+$$p$ collisions allow us to directly probe 
gluons and thereby constrain $\Delta G$. Experimentally, the observable 
used for this analysis is the double longitudinal spin asymmetry
\begin{equation}
A_{LL}\equiv \frac{\sigma^{++}-\sigma^{+-}}{\sigma^{++}+\sigma^{+-}},
\label{alldef}
\end{equation}
where $\sigma^{++ (+-)}$ is the cross section with same (opposite) sign 
helicity states of the incoming protons. In pQCD, the numerator in 
Eq.~\ref{alldef} is proportional to $\Sigma_{ij}\Delta f_i \otimes \Delta 
f_j \otimes \Delta\hat{\sigma}_{ij} \otimes D^{h}_{k}$, where $\Delta f_{i 
(j)}$ are the helicity-dependent PDFs of parton type $i$ ($j$) and depend 
on the partonic momentum fraction $x$ and on the factorization scale 
$\mu_F$, $\Delta \hat{\sigma} = \hat{\sigma}^{++}-\hat{\sigma}^{+-}$ is 
the polarized partonic cross section that depends on the renormalization 
scale $\mu_R$ and $D^{h}_{k}$ is the FF for the hadronization of outgoing 
parton $k$, which is a function of the fragmentation scale $\mu_{F'}$. For 
particle production involving $q$-$g$ or $g$-$g$ partonic scattering 
processes, information on $\Delta G$ is encoded in Eq.~\ref{alldef}.

The $A_{LL}$ of various singly inclusive (single-particle/jet) and doubly 
inclusive (two-particle/jet) production are measured at RHIC. The recent 
addition of single-inclusive $\pi^0$ 
\cite{Adare:2007dg,Adare:2008aa,Adare:2008qb,Adare:2014hsq} and jet 
$A_{LL}$ \cite{Abelev:2006uq,Adamczyk:2012qj} measurements to a global 
analysis by de~Florian, et al. (DSSV) 
\cite{deFlorian:2008mr,deFlorian:2009vb,deFlorian:2014yva} is starting to 
impose significant constraints on $\Delta G$. However, improving 
systematic uncertainties at low $p_T$ arising from experimental as well as 
theoretical sources remains a challenge. In addition, the complex nature 
of extracting PDFs and FFs by fitting data sets from various experiments 
demands independent measurements through the production of different 
final-state particles covering a wide kinematic range.

Single inclusive midrapidity charged pion production for 
$5<p_T<12$~GeV/$c$ from 200~GeV $p$$+$$p$ collisions is dominated by the 
$q$-$g$ hard process~\cite{Adare:2010cy} and the polarized partonic cross 
sections for all the relevant hard processes are positive. The signs of 
$\Delta u$ and $\Delta d$ are known to be positive and negative, 
respectively (see e.g.~\cite{deFlorian:2008mr}). Also, $u$-quarks 
preferentially fragment into $\pi^+$ and $d$-quarks into $\pi^-$. 
Consequently, the double longitudinal spin asymmetries for the three $\pi$ 
meson species should be ordered as 
$A^{\pi^+}_{LL}>A^{\pi^0}_{LL}>A^{\pi^-}_{LL}$ for a positive $\Delta G$ 
and vice versa for a negative $\Delta G$ in leading order pQCD.

A more quantitative interpretation requires the inclusion of such data 
into a global fit using the next-to-leading order (NLO) pQCD framework. 
The midrapidity production of charged pions with $5<p_T<12$~GeV/$c$ at 
$\sqrt{s}=200$~GeV covers the kinematic range of $0.03\lesssim x \lesssim 
0.16$. The relevant ingredients for a global analysis are available: 
unpolarized quark and gluon PDFs, polarized quark PDFs, charge-separated 
unpolarized FFs \cite{deFlorian:2007aj} and hard scattering cross sections 
at NLO. The invariant differential cross sections for $\pi^+$ and $\pi^-$ 
as a function of $p_T$ can be used to check the validity of the NLO pQCD 
calculation as well as the PDFs and FFs adopted for the global analysis on 
$\Delta G$.

The double-spin asymmetry $A_{LL}$ for inclusive charged pion production 
is measured as
\begin{equation}
A_{LL} = \frac{1}{\langle P_B\cdot P_Y\rangle}\frac{N^{++}-R\cdot 
N^{+-}}{N^{++}+R\cdot N^{+-}} , R=\frac{L^{++}}{L^{+-}}
\end{equation}
where $N$ is the number of charged pions and $L$ is the luminosity for a 
given helicity combination. The notation $++$ ($+-$) follows the same 
convention as in Eq.~\ref{alldef}. The polarizations of the two 
counter-circulating RHIC beams are denoted as $P_B$ and $P_Y$ and for 2009 
were 0.56 and 0.55, respectively. The luminosity-weighted beam 
polarization product $\langle P_B P_Y\rangle$, important for $A_{LL}$, was 
0.31 with a global relative scale uncertainty of 6.5\% on the product.  
An additional uncertainty based on the precision with which we can 
determine the degree of longitudinal polarization in the 
collision~\cite{Adare:2014hsq} must be included, leading to a total 
relative scale uncertainty of $^{+7.0\%}_{-7.7\%}$.

The relative luminosity, R, between the sampled luminosities for the 
different helicities is determined from the yield of BBC triggered events 
on a fill-by-fill basis.  The systematic uncertainty on relative 
luminosity is determined by comparing to the yield of ZDC 
triggers~\cite{Adare:2014hsq}, and was found in 2009 to be 
$1.4\times10^{-3}$.

Beyond the systematic uncertainties from polarization and relative 
luminosity, the dominant systematic uncertainty on the asymmetries are 
from tracks misidentified as charged pions. The size of the possible 
asymmetry from this background was determined to be $\sim10^{-4}$. The 
determination was performed by calculating the spin asymmetries of charged 
tracks found on the opposite side of the detector after reflecting a 
charged pion track around the vertical plane. The tracks in this sample 
were required to have an otherwise valid HBD cluster with charge between 1 
and 20. It was confirmed the charge distribution of the actual clusters 
associated with such tracks follows the characteristic exponential decay 
of scintillation photons~\cite{Anderson:2011jw}, ensuring that they are 
indeed a good sample of misidentified pions.

\begin{figure}[htb]
\includegraphics[width=1.0\linewidth]{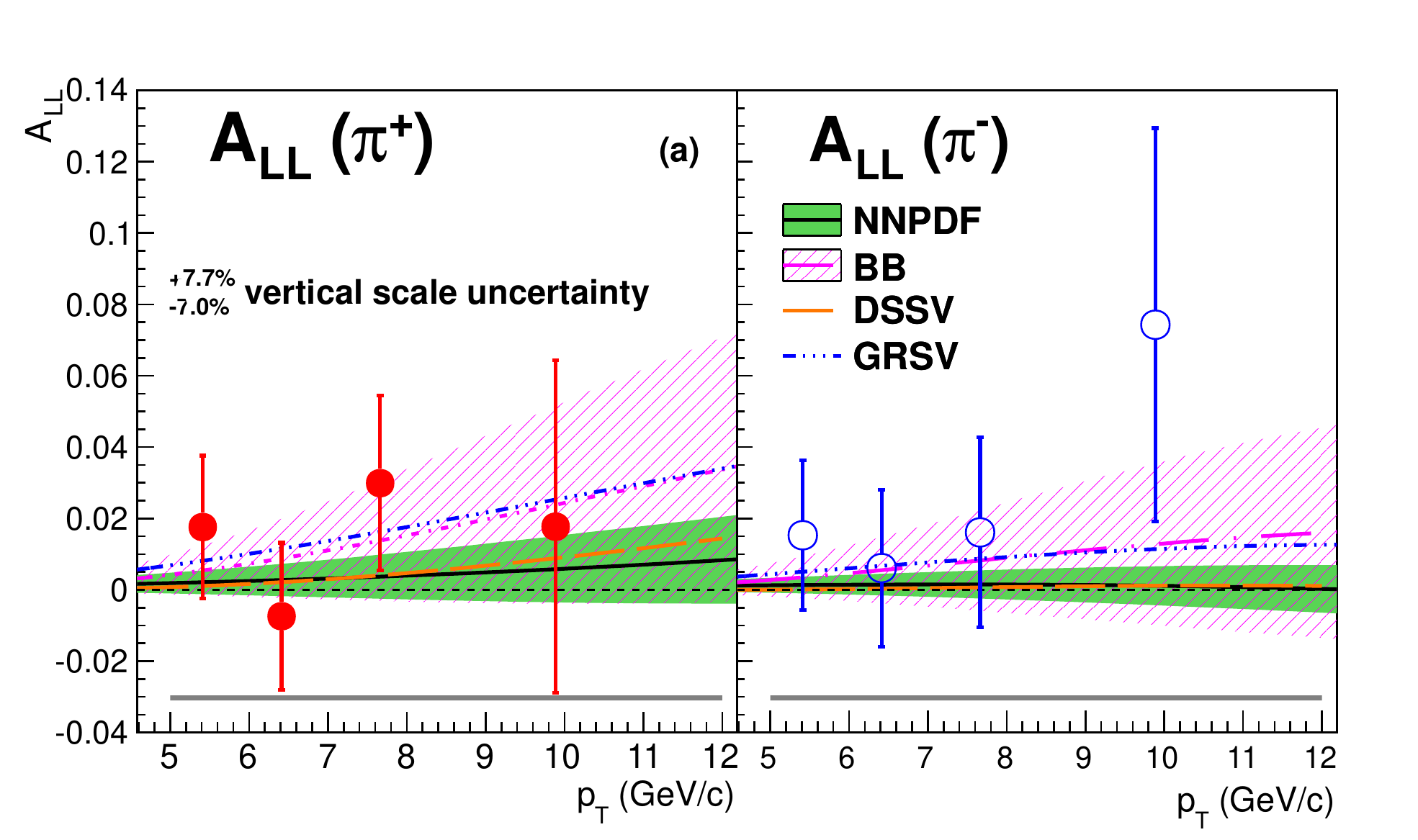}
\caption{\label{asympipm}  (color online)
The double-helicity asymmetries for (a) $\pi^+$ and (b) $\pi^-$ produced 
in $p$$+$$p$ collisions at $\sqrt{s}=200$~GeV for $|\eta|<0.35$.  The 
$p_T$-correlated systematic uncertainty from the relative luminosity is 
shown in gray.  The $^{+7.0\%}_{-7.7\%}$ scaling uncertainty is from beam 
polarization.}
\end{figure}

\begin{table}[tbh]
\caption{Double-helicity asymmetries and statistical uncertainties for 
$\pi^+$ and $\pi^-$ hadrons.  The primary systematic uncertainties, which 
are fully correlated between points, are $1.4\times10^{-3}$ from relative 
luminosity and a $^{+7.0\%}_{-7.7\%}$ scaling uncertainty from beam 
polarization.}
\begin{ruledtabular} \begin{tabular}{cccccccc}
&
&& \multicolumn{2}{c}{$\pi^+$}
&& \multicolumn{2}{c}{$\pi^-$} \\
  $p_T$ bin  & $\langle p_T\rangle$
&& $A_{LL}$ & $STAT$
&& $A_{LL}$ & $STAT$ \\
(GeV/$c$) & (GeV/$c$)
&&  &
&&  & \\
\hline
5--6    &  5.41  &&  0.018  &  0.020  &&  0.015  &  0.021  \\
6--7    &  6.41  && -0.007  &  0.021  &&  0.006  &  0.022  \\
7--9    &  7.66  &&  0.030  &  0.025  &&  0.016  &  0.027  \\
9--12  &  9.89  &&  0.018  &  0.047  &&  0.074  &  0.055  \\
\end{tabular} \end{ruledtabular}
\label{t:pi0_final_datatable}
\end{table}

Charge-separated pion $A_{LL}$ measurements are shown in 
Fig.~\ref{asympipm}. As statistical errors dominate the uncertainties, 
point-to-point systematic errors are not plotted.

\section{Discussion}
\label{sec:discussion}
\subsection{Cross Sections and Charge Ratio}

In Fig.~\ref{xsecpipm}, the charged pion cross sections are compared to 
pQCD calculations~\cite{marco} which were performed at NLO using CTEQ6.5 
unpolarized PDFs~\cite{Tung:2006tb} and 
DSS~\cite{deFlorian:2007aj,Epele:2012vg} and AKK~\cite{Albino:2008fy} FFs. 
In the bottom panel of each figure, the relative difference between data 
and theory is also shown for the DSS FFs.  The absolute normalization 
uncertainty of 9.6\% is not included in the systematic errors shown in 
boxes.  A standard technique for determining theoretical uncertainties is 
to vary the renormalization, factorization, and fragmentation theory 
scales by a set factor, in this case 2.  Here the three scales, denoted 
$\mu$, are set to be equal, and are varied from $\mu = 0.5p_T$ to $\mu = 
2p_T$.  The pion measurements fall within $\sim \pm20$\% of the pQCD 
calculation at a scale of $\mu = p_T$, well within the much larger range 
defined by varying the scales.  Note that no other uncertainties on the 
pQCD calculations such as those due to PDF uncertainties or FF 
uncertainties are included.

The measured ratio of $\pi^-$-to-$\pi^+$ production is shown in 
Fig.~\ref{ratio}.  At low $p_T$ where $\pi^{\pm}$ production is dominated 
by $g$-$g$ scattering, the measured charge ratio of pion production is 
close to 1. In contrast, $q$-$g$ scattering dominates $\pi^{\pm}$ 
production at higher $p_T$ and the ratio starts to deviate from 1 due to 
the valence quark content of the proton. Any hadrons that fragment 
preferentially from $u$-quarks are enhanced with respect to hadrons that 
fragment preferentially from $d$-quarks, leading to an increase of 
positive with respect to negative pions.  As presented in 
Fig.~\ref{ratio}, the pQCD calculations of the $\pi^-$-to-$\pi^+$ ratio 
lie above the measured ratio for $p_T$$>$$\sim$2--3~GeV/$c$ by as much as 
$\sim$20\% for both the DSS~\cite{deFlorian:2007aj} and 
AKK~\cite{Albino:2008fy} FFs.  We note that the calculated midrapidity 
$\bar{p}$-to-$p$ ratio using DSS FFs~\cite{deFlorian:2007hc} exceeds the 
ratio measured by PHENIX by 20\%--40\%~\cite{Adare:2011vy}.  The 
theoretical curves shown are simply obtained by dividing the two cross 
section calculations for each of the different scales.  The range of 
calculated ratios indicates disagreement with the measured ratio presented 
in this paper as well as that from STAR~\cite{Agakishiev:2011dc}; the 
discrepancy implies that other sources of uncertainty in the ratio 
calculation need to be investigated.  Uncertainties for the DSS FFs are 
estimated in~\cite{Epele:2012vg}, but a calculation taking into account 
these uncertainties and which components cancel or do not cancel in the 
ratio is not currently available.  Given extensive work in recent years to 
develop Lagrange multiplier and Hessian techniques to assess uncertainties 
on PDFs and FFs following early work by the CTEQ 
collaboration~\cite{Stump:2001gu}\cite{Pumplin:2001ct}, the community 
should soon reach a point where it is standard to propagate PDF and FF 
uncertainties fully for calculations of observables.  We note that much of 
the data used in the pion FFs comes from $e^+e^-$ annihilation, which 
provides very little sensitivity to the quark flavor dependence. Only a 
limited amount of charge-separated data from SIDIS and the BRAHMS 
experiment at RHIC is incorporated in the DSS FFs.  The AKK FFs include 
charge-separated data for pions from BRAHMS and an earlier STAR 
measurement~\cite{Adams:2006nd}, but no SIDIS data.  The DSS and AKK FFs 
include $\pi^0$ data from PHENIX and STAR, which improves constraints on 
gluon FFs to pions but does not provide sensitivity to quark flavor due to 
the zero isospin of the neutral pion.  Furthermore, the SIDIS and $e^+e^-$ 
data included in the fit are at a lower average fraction of the jet 
momentum ($z$) than the RHIC measurements, leading to weaker data-based 
constraints on the FFs in the $z$ range most relevant to RHIC. The 
relevant $z$ range covered by the present measurements is 
$\sim$~0.45--0.71, determined by extracting the $z$ distribution for the 
pQCD calculations shown in this paper for $5 < p_T < 13$~GeV/$c$ using the 
DSS and AKK FFs independently. The calculations using the two FFs showed 
consistent results.

\begin{table}[tbh]
\caption{Ratio of charged pion cross section, as shown in Fig.~\ref{ratio}.}
\begin{ruledtabular} \begin{tabular}{ccccc}
$\langle p_T\rangle$
&& Ratio & STAT & SYST  \\
(GeV/$c$)
&&  &  & \\
\hline
5.39    &&  0.850  &  0.035  &  0.027  \\
6.39    &&  0.858  &  0.037  &  0.027  \\
7.41    &&  0.821  &  0.052  &  0.026  \\
8.44    &&  0.798  &  0.075  &  0.026  \\
9.71    &&  0.733  &  0.083  &  0.023  \\
11.76  &&  0.74    &  0.16    &  0.022
\end{tabular} \end{ruledtabular}
\label{t:pi0_final_datatable}
\end{table}

We note that a similar discrepancy exists between the calculated and 
measured $\eta$-to-$\pi^0$ ratio at midrapidity in $p$$+$$p$ 
collisions~\cite{Adare:2010cy, Abelev:2012cn}.  Both the $\eta$ 
FFs~\cite{Aidala:2010bn} and $\pi^0$ FFs~\cite{deFlorian:2007aj} included 
PHENIX data in the fits, which helped to constrain the fragmentation from 
gluons in particular.  However, with the $\eta$ and pion FF 
parameterizations performed independently, the correlation of the 
normalization uncertainty on the PHENIX pion and $\eta$ cross sections was 
not taken into account, and the normalization was scaled within the 
uncertainty in opposite directions for the two measurements to 
minimize the $\chi^2$ when the fits were performed along with other world 
data.  We propose that future FF parameterizations include direct fits of 
particle production ratios in cases where data are available.  We expect 
that fitting ratios directly could significantly improve constraints on 
FFs because of cancellations in systematic uncertainties both in the 
measured data and in the calculations.  Improved knowledge of FFs can in 
turn improve extractions of helicity PDFs as well as other nonperturbative 
functions related to hadron structure.

\subsection{Helicity Asymmetries}

In Fig.~\ref{asympipm}, our charged pion $A_{LL}$ results are compared 
with three different expectations based on different global analyses, or 
fits, of helicity PDFs to world polarized data.  The 
Bl\"{u}mlein-B\"{o}ttcher (BB) fit~\cite{Blumlein:2010rn} and the 
de~Florian, et~al. (DSSV) fit~\cite{deFlorian:2009vb} use a specified 
functional form to describe the helicity PDFs, while 
NNPDF~\cite{Ball:2013lla,Ball:2012cx} use a neural network framework 
without a specified functional form, allowing for additional freedom in 
the fit.  Both BB and NNPDF used only polarized DIS data for their 
constraint; the DSSV fit also includes SIDIS data as well as RHIC data for 
$\pi^0$ and jet $A_{LL}$, which were found to significantly constrain 
$\Delta G$ in the intermediate $x$ range, 0.05--0.2.  NNPDF recently 
released an updated helicity PDF fit to include RHIC jet and $W$ boson 
data~\cite{Nocera:2014gqa}, but this new fit has not been used for the 
calculations shown in Fig.~\ref{asympipm}.  The generally larger 
asymmetries predicted for positive pions than negative pions reflect a 
$\Delta G$ that is positive in the fits that are used for these 
calculations.  One must take care in directly comparing our data with 
these expectations, as the DSS FFs have been used to calculate the 
$A_{LL}$ expectations. As discussed above, the accuracy of any FF when 
comparing positive to negative charged pions needs further study.

\begin{figure}[htb]
\includegraphics[width=1.0\linewidth]{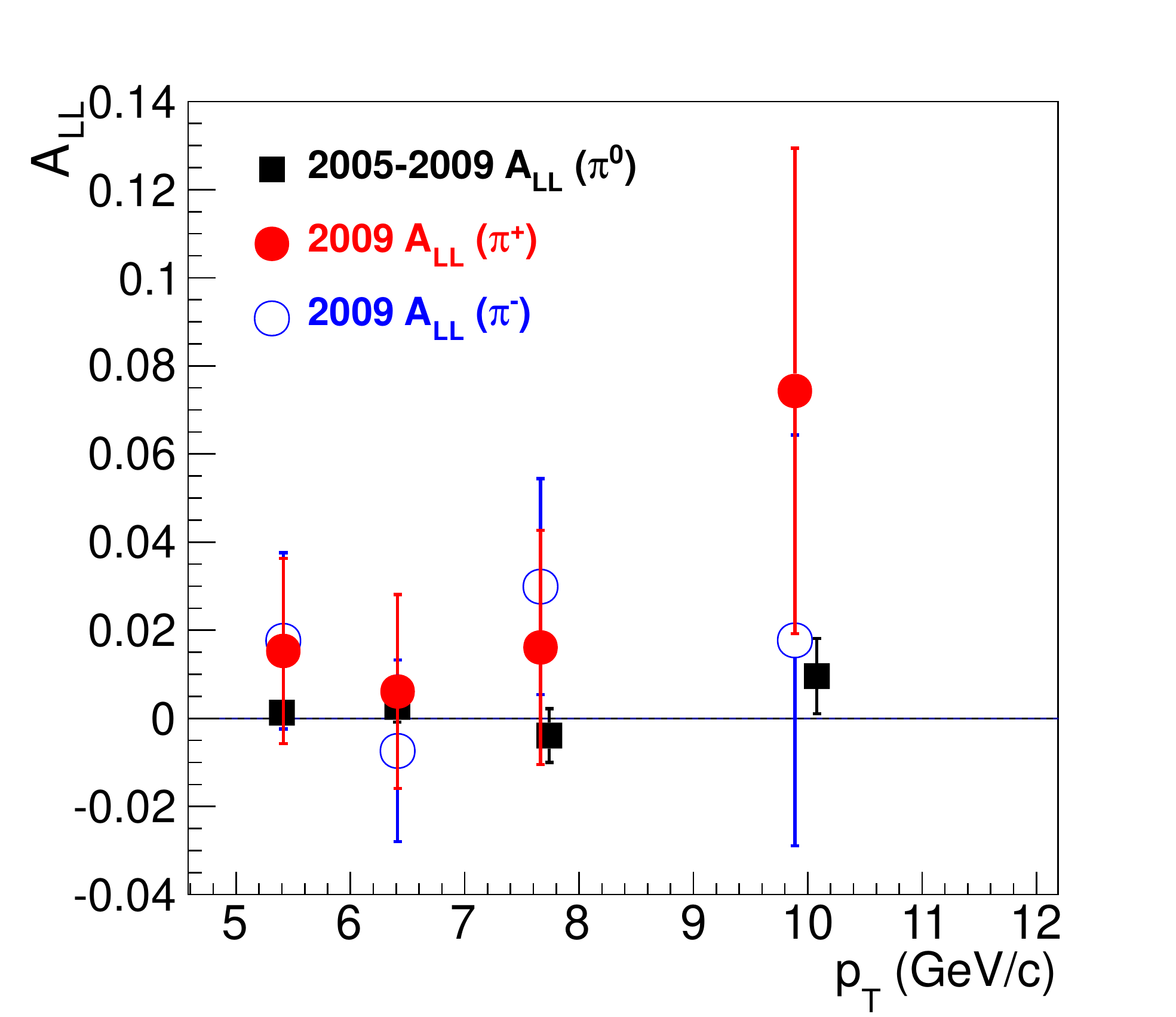}
\caption{\label{asympi}  (color online)
Comparisons of double-helicity asymmetries for midrapidity positive, 
negative, and neutral pion production in $p$$+$$p$ collisions at 
$\sqrt{s}=200$~GeV measured by PHENIX.  The neutral pion data are 
from~\cite{Adare:2014hsq}.}
\end{figure}

In Fig.~\ref{asympi}, we compare our charged pion $A_{LL}$ results with 
our previously published $\pi^0$ $A_{LL}$ results.  By comparing the 
ordering of the charged and neutral pions, one could get information on 
the sign of $\Delta G$ independent of any FF assumptions.  However, due to 
the lack of a dedicated hadron trigger in PHENIX, the statistical 
precision of the charged pion data is limited, and does not allow for 
clear sign determination with the current data.  In future global 
analyses, the inclusion of these data should enhance sensitivity to the 
sign of the gluon polarization.

\section{Summary}
\label{sec:summary}

In summary, the invariant cross sections, ratio, and double-helicity 
asymmetries of charge-separated positive and negative pions produced at 
midrapidity in $\sqrt{s} = 200$~GeV $p$$+$$p$ collisions have been measured.  
The $p_T$ range of the cross section measurements is from 5--13~GeV/$c$; 
that of the asymmetries is from 5--12~GeV/$c$. The separate positive and 
negative cross section measurements are consistent with NLO pQCD 
calculations within a large theoretical scale uncertainty and fall within 
$\sim \pm20$\% of the calculation at $\mu = p_T$ over the range presently 
measured.  These charged pion cross section results, when included in FF 
fits, should improve predictions for future measurements.  The NLO pQCD 
predictions for the ratio of negative to positive pion cross sections lie 
above the measurement by as much as 20\%.  This 20\% difference does not 
fall within the range of ratios calculated using different choices of 
scale, the only uncertainty presently available on the ratio calculation, 
and indicates that other sources of systematic uncertainty on the ratio 
calculation need to be investigated.  Inclusion of neutral pion data from 
RHIC in existing FF parameterizations has significantly improved 
constraints on the gluon-to-pion FF.  Future FF fits which incorporate the 
present data, particularly the ratio data, will especially improve 
constraints on the flavor dependence of quark FFs to pions.  To 
advance FF parameterizations more generally, we recommend that the 
phenomenology community move toward fitting measured particle ratios 
directly in cases where data are available because of the reduced 
uncertainties on both the measured and calculated quantities.  In the 
future the charge-separated asymmetries with improved FFs in hand should 
be included in an updated global analysis of helicity PDFs.  These data 
can increase sensitivity to the sign information of $\Delta G$ in future 
pQCD helicity PDF fits.

\section*{ACKNOWLEDGMENTS}     

We thank the staff of the Collider-Accelerator and Physics
Departments at Brookhaven National Laboratory and the staff of
the other PHENIX participating institutions for their vital
contributions.  We acknowledge support from the
Office of Nuclear Physics in the
Office of Science of the Department of Energy, the
National Science Foundation, Abilene Christian University
Research Council, Research Foundation of SUNY, and Dean of the
College of Arts and Sciences, Vanderbilt University (U.S.A),
Ministry of Education, Culture, Sports, Science, and Technology
and the Japan Society for the Promotion of Science (Japan),
Conselho Nacional de Desenvolvimento Cient\'{\i}fico e
Tecnol{\'o}gico and Funda\c c{\~a}o de Amparo {\`a} Pesquisa do
Estado de S{\~a}o Paulo (Brazil),
Natural Science Foundation of China (P.~R.~China),
Ministry of Education, Youth and Sports (Czech Republic),
Centre National de la Recherche Scientifique, Commissariat
{\`a} l'{\'E}nergie Atomique, and Institut National de Physique
Nucl{\'e}aire et de Physique des Particules (France),
Bundesministerium f\"ur Bildung und Forschung, Deutscher
Akademischer Austausch Dienst, and Alexander von Humboldt Stiftung 
(Germany),
Hungarian National Science Fund, OTKA (Hungary),
Department of Atomic Energy and Department of Science and Technology 
(India),
Israel Science Foundation (Israel),
Basic Science Research Program through NRF of the Ministry of Education 
(Korea),
Physics Department, Lahore University of Management Sciences (Pakistan),
Ministry of Education and Science, Russian Academy of Sciences,
Federal Agency of Atomic Energy (Russia),
VR and Wallenberg Foundation (Sweden),
the U.S. Civilian Research and Development Foundation for the
Independent States of the Former Soviet Union,
the Hungarian American Enterprise Scholarship Fund,
and the US-Israel Binational Science Foundation.


%

%
 
\end{document}